\documentclass{aastex}
\usepackage{spr-astr-addons}
\usepackage{epsfig}

\shorttitle{An Interesting Paper}
\shortauthors{Enthusiasticus et al.}
\begin{document}


\title{Reddening law and interstellar dust properties along Magellanic sight-lines}


\author{Fr\'ed\'eric Zagury\altaffilmark{1} }
\affil{Institut Louis de Broglie, 23 rue Marsoulan, 75012 Paris, France}


\altaffiltext{1}{Visiting Astronomer, Konkoly Observatory, Budapest.}
%

\begin{abstract}
This study establishes that SMC, LMC and Milky Way extinction curves obey the same extinction law which depends on the $2200\,\rm\AA$ bump size and one parameter, and generalizes  the Cardelli, Clayton \& Mathis (1989) relationship.
This suggests that extinction in all three galaxies is of the same nature.
The role of linear reddening laws over all the visible/UV wavelength range, particularly important in the SMC but also present in the LMC and in the Milky Way, is also highlighted and discussed.
\end{abstract}



%
%
%

\section{Introduction} \label{intro}
Extinction curves in the  \objectname{Small Magellanic Cloud} (\objectname{SMC}) have a striking shape close to linearity in wavenumber 
$1/\lambda$, over all the  visible and UV wavelength range \citep{prevot84, g03, cartledge05}. 
Taking into account the forward-scattering properties of interstellar dust, both in the visible and in the UV  \citep{henyey41,z00a}, this kind of curves, well known in Aeronomy, accepts a simple interpretation:
extinction in these directions must be due to a power law size distribution of interstellar large (with size comparable or larger than the wavelength) grains.
No information on the precise composition of the grains  can be deduced from such extinction curves however: aerosols in the earth atmosphere for instance produce a similar extinction law.

The  \citet{prevot84} conclusion that linearity and the absence of a $2200\,\rm\AA$ bump characterize SMC extinction is not totally right:
AzV456 has a bump \citep{g03}, and the extinction curve of  AzV398 departs from linearity in the far-UV (Sect.~\ref{sb}).
In fact, recent studies  \citep{g03, cartledge05} show that SMC and  \objectname{Large Magellanic Cloud} ( \objectname{LMC}) extinction curves share  the same general features one finds in the solar neighboorhood: a linear law in the visible, the $2200\,\rm\AA$ bump, and a far-UV rise.
But, for an equal reddening (measured by $E(B-V)$), Magellanic extinction curves will in most cases
differ from average Galactic ones by
a less important $2200\,\rm\AA$ bump and 
a steeper far-UV rise \citep{cartledge05}: SMC linear curves constitute the extreme case 
where the bump disappears and the far-UV rise prolongs the linear 
visible extinction law.
Conversely, Magellanic-like extinction curves are found in the Milky Way \citep{clayton00},  Galactic directions with no bump exist at relatively high reddenings (planetary nebulae NGC6905 and NGC5189, with $E(B-V)\sim 0.6$), and linear extinction curves are found in the Galaxy along directions of low column density (Sect.~\ref{bvn} and \ref{nb} this paper, and \citet{z01}).

In the matter of the variations of the observed properties of interstellar extinction in and outside the Galaxy, one may therefore consider that the differences between Magellanic and Galactic extinctions are not so clear.
The fact that Magellanic-like extinction curves are not fitted by the one parameter free, empirical,  \citet{ccm89}
(CCM) relationship can not be used as a proof, as it has been argued \citep{g03,cartledge05}, of differences in grain compositions between Magellanic-like and average Galactic-like sight-lines:
its failure to reproduce linear SMC extinction curves (Sect.~\ref{ccm}) may simply show its limits,  awaiting for a more general, physically meaningful, understanding of interstellar extinction.
In this respect, the obtaining of a general, common fit to Magellanic and Galactic extinction curves, suggested in \citet{g03}, could represent a major step in the comprehension of the similarities and differences of interstellar extinction between the three galaxies.
The number of free parameters of such a fit [in \citet{g03} it is suggested that at least two parameters, $R_V$ and HI column density, are necessary to fix the extinction curve in any direction] will not only determine the degree of similarity between extinction laws in these galaxies but may also shed light on the dependencies of interstellar extinction in the Universe.

In a more theoretical perspective, the existence of a linear interstellar extinction law over the whole visible and UV wavelength range, its importance, and its relationship to the traditional CCM extinction law, have been poorly investigated, mainly because extinction curves in the solar neighborhood are mostly CCM-like.
This raises significant questions.
It first implies that the near-UV break and the important UV flattening-out in the extinction law of large grains assumed by all grain models (see for instance fig.~2 in \citet{desert90})  is not that obvious.
It further means that, in the Magellanic Clouds as well as in our Galaxy, either CCM-like and linear extinction laws coexist in the interstellar medium, or,  there is a progressive passage, though at variable critical values of the reddening, from linearity to the more complex CCM extinction curves.

This missing link between linear and CCM extinction laws, the possible role of a linear extinction law in high reddening directions, and the fundamental question of the differences and similarities between Magellanic and Galactic dusts, will be my main preoccupations in this article.
\section{Data} 
   \label{data}
This paper relies on a comparison of the spectral energy distributions (SEDs) of Magellanic stars in the visible and in the UV.
The UV wavelength domain is paradoxically the most documented one thanks to the International Ultraviolet Explorer (IUE) database which covers a large part of the sky.
No equivalent database exist for the visible and for the near-IR, where SEDs of individual stars are finally poorly known.
Most studies on visible extinction rely on photometry, the main indication of a reddening difference between two stars of same spectral type being  $\Delta(B-V)$, the difference of $B-V$ between the stars.
It is clear however that $\Delta (B-V)$ is not as good an indicator of reddening as a direct comparison of the visble SEDs can be.

The forty one stars used in this paper are those of \citet{g03} (table~2) with three exceptions: 
AzV23, for which no IUE spectrum was found in the archives of the Vilspa ESA centre, 
SK-67~279, whose only LWR available spectrum is not reliable, and Sk-69~206.
The latter has an abnormally low $E(B-V)\sim 0.25$ with respect  to the important $2200\,\rm\AA$ bump in its UV spectrum.
The UV spectrum of Sk-69~206 is exactly proportional with that of the neighbor star Sk-69~210, which has $E(B-V)\sim 0.53$.
A bias in the visible photometry, or the possibility that $B-V$ is not representative of the average slope of the visible spectrum for this star could explain the discrepancy; a comparison of Sk-69~206's visible spectrum to that of   Sk-69~210 or of an unreddened star of same spectral type would be the most appropriate way to understand the visible extinction law in this direction.
 
SEDs ratios ($F_{\star 1}/F_{\star 2}$) will be used (rather than magnitude differences) and I call  'reduced spectrum' of a reddened star the ratio $F_\star/F_0$ of the star spectrum to that of a non (or slightly) reddened one of same spectral type. 
The difference with traditional extinction curves (given by $-2.5\log(F_\star/F_0)+Cte$) is minimal when reduced spectra are plotted on a logarithmic scale; reduced spectra have the advantage of being the direct expression of the observations, and their use was found to be more convenient for this work.
\subsection{Visible photometry} \label{vis}
\begin{table*}
\caption[]{SMC data}		
       \[
    \begin{tabular}{lccccccc}
\hline
star &$B-V$$^{(1)}$ & \multicolumn{3}{c}{spectral type} &$E(B-V)$$^{(2)}$&B$^{(3)}$ \\
&  & Vis$^{(4)}$&UV$^{(5)}$&type$^{(6)}$  && \\
\hline
AzV18 & $0.041\pm 0.006$ &B1Ia & B3Ia &B0-B1.9 &0.23/0.17 &N \\
AzV70  & $-0.154\pm 0.013$  & O9.5Iw & O9Ia & O &0.12/0.13 & N\\
AzV214 & $0.038\pm 0.007$ &  B3Iab& B2Ia & B0-B1.9 & 0.17/0.21 & N\\
AzV289& $-0.118\pm 0.009$ &   B0I& O9Ia & O & 0.12/0.16 &N \\
AzV380  & $-0.109\pm 0.010$ & B0.5Ia &B0Ia & B0-B1.9 & 0.11/0.13 &N \\
AzV398  & $0.100\pm 0.022$ & O9.7Ia & O9Ia & O &0.37/0.38 & N\\
AzV404 & $-0.098\pm 0.006$ & B2Ia &B3Ia &B0-B1.9  &0.07/0.03  & N\\
AzV456  & $0.109\pm 0.009$ & O9.7Ib &O8II & B0-B1.9 &0.38/0.43 & W\\
AzV462& $-0.126\pm 0.012$ &   B1Ia &B2Ia & B0-B1.9 & 0.05/0.04 & W\\
\hline
\end{tabular} 
 \]
\begin{list}{}{}
\item[$(1)$] From \citet{g03}.
\item[$(2)$] Indicative reddenings derived from columns~4 and 5, and from $(B-V)_0$ in \citet{fitzgerald70}.
\item[$(3)$]'N' for no $2200\,\rm\AA$ bump, 'W' for a weak bump, 'L' for a large bump.
\item[$(4)$]From studies in the visible (Sect.~\ref{vis}).
\item[$(5)$]From \citet{snb97,snb99}.
\item[$(6)$]Adopted types (Sect.~\ref{uv}). 
\end{list}
\label{tbl:smc}
\end{table*}
$B-V$ photometry in Tables~\ref{tbl:smc} and \ref{tbl:lmc} is essentially the same as in \citet{g03}.

I found two sets  of photometric data for SMC stars, one which can be retrieved from SIMBAD (original sources in \citet{azzopardi75}, \citet{ardeberg77}, or \citet{nicolet78}) and the more recent one of  \citet{g03}.
The agreement is generally good except for AzV214 and AzV462 (for which SIMBAD gives respectively $B-V=0.07$ and $-0.17\pm 0.02$).
\citet{g03} $B-V$ values have been adopted here (column~2 of Table~\ref{tbl:smc}).
Spectral information for SMC stars are from either \citet{azzopardi82} or \citet{garmany87}.

LMC photometry (Table~\ref{tbl:lmc}) was  retrieved from SIMBAD,  original sources being in either \citet{ardeberg72}, \citet{isserstedt75,isserstedt79,isserstedt82}, or \citet{nicolet78}.
They generally result from three to four  measurements, and are given with an uncertainty of $\pm0.02$~mag.
A few stars with more precise photometry were found at the Lausanne GCPD database (http://obswww.unige.ch/gcpd).
LMC spectral classification from visible wavelengths (third column of Table~\ref{tbl:lmc}) is from either \citet{rousseau78} or \citet{fitz88}.

Indicative $E(B-V)$ reddenings are derived at the end of Tables~\ref{tbl:smc} and \ref{tbl:lmc} from the visible photometry, the visible/UV spectral types, and from the intrinsic \citet{fitzgerald70} colors.
\begin{table*}
\caption[]{LMC data (see Table~\ref{tbl:smc} for columns meaning)}		
       \[
    \begin{tabular}{lcccccc}
\hline
star & $B-V$ & \multicolumn{3}{c}{spectral type} &$E(B-V)$&B \\
&  & Vis&UV&type  & &\\
\hline
Sk-65 15  & $-0.11$  & B1Ia &- & B2-B4 & $0.08/-$ &  N\\
Sk-65 63  & $-0.16$  & O9.7I& - &O&  0.1/- &N \\
Sk-66 19 & $0.12$ &B4I & B0Ia  & B0-B1.9 &  0.23/0.36 & W\\
Sk-66 35  & $-0.077\pm 0.005$ (4) & B1Ia& -&B2-B4  &  0.11/- & N\\
Sk-66 88 & $0.20$  & B2& -  &B2-B4  &  0.37/- & W\\
Sk-66 106 & $-0.08$  & B2Ia &- & B2-B4 &  0.09/- &N \\
Sk-66 118  & $-0.05$ & B2Ia&- & B2-B4 &  0.12/- & N\\
Sk-66 169  & $-0.13$  & O9.7Ia& O9Ia &O  &  0.13/0.15 &N \\
Sk-67 2 &  $0.099\pm 0.009$  & B1.5Ia& B2Ia & B2-B4 &  0.28/0.27 &W \\
Sk-67 5  & $-0.115\pm 0.011$ & O9.7Ib & B0Ia & B0-B1.9 & 0.15/0.12 &N \\
Sk-67 36  & $-0.08$  & B2.5Ia&B3Ia & B2-B4  & 0.07/0.09 & N\\
Sk-67 78  & $-0.04$  & B3Ia &- & B2-B4 & 0.07/- &N \\
Sk-67 100  & $-0.09$  & B1Ia& - & B2-B4 &  0.1/- &N \\
Sk-67 168 & $-0.17$ & O8Ia &- & O &  0.12/- &N \\
Sk-67 228 & $-0.05$  & B2Ia &B2Ia & B2-B4 &  0.12/0.12 &N \\
Sk-67 256 & $-0.08$ & B1Ia &- &  B2-B4 & 0.11/- & N\\
Sk-68 23  & $0.22$  & OB &B3Ib & B2-B4 &  0.5/0.34 &L \\
Sk-68 26  & $0.13$  & B8I &B3Ia & B2-B4 &  0.14/0.26 &W \\
Sk-68 40 & $-0.07$  & B2.5Ia& - & B2-B4 &  0.08/- &N \\
Sk-68 41& $-0.11$  & B0.5Ia &-& B0-B1.9  & 0.08/- &N \\
Sk-68 129  & $0.03$  & B0.5&- & B0-B1.9 &  0.25/- &W \\
Sk-68 140  & $0.09$ & B0& - & B0-B1.9 &  0.33/- &W \\
Sk-68 155  & $0.03$  & B0.5 &- & O &  0.25/- &W \\
Sk-69 108  & $0.27$  & B3I & B4Ia & B2-B4 &  0.50/0.50 & L \\
Sk-69 210  & $0.36$  & B1.5 &- & B2-B4  & 0.53/- &L \\
Sk-69 213  & $0.10$ & B1&- & B2-B4 &  0.29/- &W \\
Sk-69 228 & $0.07$ & OB&B2Ia & B2-B4  & 0.34/0.24 &N \\
Sk-69 256 & $0.03$  & B0.5 & -& B2-B4 &  0.25/- &N \\
Sk-69 265 & $0.12$ & B3I& B3Ia & B2-B4 &  0.25/0.25 &W \\
Sk-69 270 & $0.14$  & B3Ia& B4Ia & B2-B4 &  0.27/0.31 &W \\
Sk-69 280 & $0.09$  & B1& B1.5Ia & B2-B4 &  0.28/0.27 &W \\
Sk-70 116 & $0.104\pm 0.005$ & B2Ia &B3Ia & B2-B4 &  0.27/0.23 & N\\
Sk-70 120  & $-0.069\pm0.008$  & B1Ia &-& B2-B4 &  0.12/- & N\\
\hline
\end{tabular} 
 \]
\label{tbl:lmc}
\end{table*}
\subsection{UV data} \label{uv}
IUE spectra were 
retrieved from the INES Archive Data Server.
\begin{figure}[t]
\resizebox{1.\columnwidth}{!}{\includegraphics{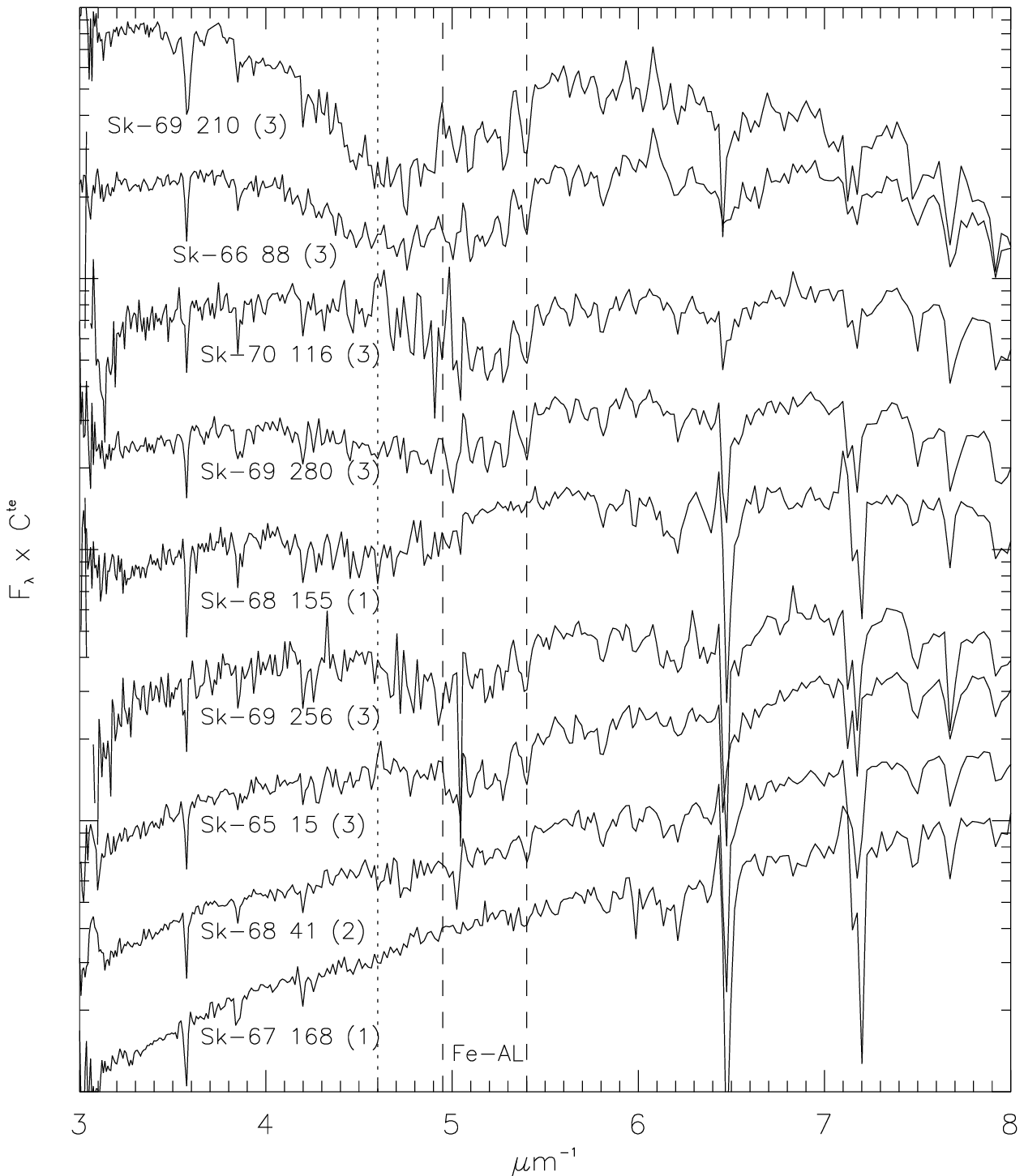}} 
\caption{UV spectra sample of LMC and SMC field stars, classified by decreasing slope, and rescaled by an arbitrary factor. Y-logarithmic axis. 
The number after each star's name stands for its type [(1)=O stars,(2)=B0-B1.9, (3)=B2-B4 (Sect.~\ref{uv})]. 
The AL/Fe region and the $2200\,\rm\AA$ ($4.6\,\rm \mu m^{-1}$) bump are indicated.
} 
\label{fig:pres0}
\end{figure}

I used the  ALIII ($1855-11863\,\rm\AA$, spectral origin) and FeIII ($1891-1988\,\rm\AA$, spectral origin with a possible interstellar contribution) prominent features to classify the stars according to their spectral type, following \citet{prevot84} in the expectation that for two stars of close spectral types the SEDs ratio will cancel the AL/Fe depression.
Stars were separated into three main types, according to the importance of the AL/Fe depression.
Compared with the Smith~Neubig~- Bruhweiler \citep{snb97,snb99} one, this classification should roughly correspond to O giants (no  ALIII-FeIII lines), B0-B2 (small depression between $4.8$ and $5.4\,\mu\rm m^{-1}$)) and B2-B4 (large depression).
Differences occasionally appeared (AzV18, AzV404, AzV456, Sk-69~280, Tables~\ref{tbl:smc} and \ref{tbl:lmc}) with the Smith~Neubig~- Bruhweiler classification,
probably due to the different methods used to determine spectral types from UV spectra: 
 the Smith~Neubig classification depends on the six order polynomial fit of the continuum, while the stars were here directly compared one to another after a correction for their average slope differences.
The spectral classification derived from UV wavelengths often differs (\citet{snb97,snb99} and Tables~\ref{tbl:smc} and \ref{tbl:lmc}) from  the visible ones.
  
Figure~\ref{fig:pres0} shows a sample of the UV spectra, with an arbitrary scaling, extracted from  the forty one stars of the sample, and classified by decreasing mean UV slope.
The y-axis is logarithmic.
An order of magnitude corresponds to the interval between two long ticks, the whole y-axis of the plot spans three orders of magnitude. 
Dashed vertical lines delimit the AL/Fe region, the central wavelength of the $2200\,\rm\AA$ bump is indicated by the dotted line.
The progressive inclination of the spectra is manifest.
From $3\,\mu\rm m^{-1}$ to $8\,\mu\rm m^{-1}$ the flux level increases by a factor of $\sim 8$ for the bottom spectra of 
Sk-67~168, while it decreases by a factor of $\sim 5$ for Sk-69~210:
the average slope of the spectrum is therefore decreased by a factor of $\sim 40$ between the two stars (the change of slope is roughly an exponential $e^{-0.7/\lambda}$, with $\lambda$ in  $\mu$m).

Modification of the average slope of the UV spectrum, from one star to another, can be due either to spectral type differences (temperature reddening), or, to interstellar extinction.
It is clear, from  the $E(B-V)$ columns of Tables~\ref{tbl:smc} and \ref{tbl:lmc}, that all stars of 
the sample, including the ones used as references in \citet{g03}, are reddened, either by Galactic dust, or, by dust in the Magellanic Clouds.
It can be presumed that the small diminution of slope between the UV spectra of Sk-67 168 (O), Sk-68~41 (B0-B1.9), or Sk-65~15 (B2-B4) is mainly a question of temperature reddening, while  differences between stars of same type (Sk-67~168 and Sk-68~155) or  important ones (Sk-67~168 and Sk-68~155 or Sk-69~210) are  more a matter of interstellar extinction.
\section{Linear reddening} \label{elin}
In the visible, a temperature difference or a difference of interstellar extinction between two stars appear as a difference of  slope between the SEDs.
In magnitudes both effects are linear in $1/\lambda$   \citep{hall37,stebbins39, divan54}.
If $(B-V)_2$ and $(B-V)_1$ are the respective colors of two stars 1 and 2, and if $\Delta_{vis}=2((B-V)_2-(B-V)_1)$, the magnitude difference at wavelength $\lambda$ (expressed in  $\mu$m, $1/\lambda_V=1.82\mu$m$^{-1}$, $1/\lambda_B=2.27\mu$m$^{-1}$) is 
\begin{equation}
m_{\lambda,2}-m_{\lambda,1}=1.06\frac{\Delta_{vis}}{\lambda}+c 
  \label{eq:dmv1},
\end{equation}
($c$ a  constant).

If the stars have same spectral type
\begin{equation}
A_{\lambda,2}-A_{\lambda,1}=1.06\Delta_{vis}(\frac{1}{\lambda}+(R_v-4))
    \label{eq:drv}
\end{equation}
and: $\Delta_{vis}=2\Delta E(B-V)$.

In terms of fluxes, Eq.~\ref{eq:dmv1} is
\begin{equation}
\frac{F_{\lambda,2}}{F_{\lambda,1}}=Ce^{\frac{\Delta_{vis}}{\lambda}}  \label{eq:dmv2}
\end{equation}
($C$ a constant).

The precision with which observed $2\Delta(B-V)$ approaches the difference of visible slopes between two stars, $\Delta_{vis}$ in Eqs.~\ref{eq:dmv1} or \ref{eq:dmv2}, depends on the precision of $B-V$ photometry, and on how good an indicator of the overall slope of each individual visible spectrum $B-V$ is.
The latter uncertainty is difficult to evaluate, and is generally neglected.
The former one should be less (Sect.~\ref{vis}) than 0.08~mag. (four times the uncertainty on $B-V$).

In the UV, as in the visible, temperature reddening is linear in $1/\lambda$.
Therefore, if the UV spectra of two stars differ, on a logarithmic scale, by a change of slope (an exponential of $1/\lambda$),  the difference of UV extinction laws between the two directions must also be linear in $1/\lambda$.
\section{Slightly reddened stars $(B-V<0)$} \label{bvn}
\begin{figure}[t]
\resizebox{1.\columnwidth}{!}{\includegraphics{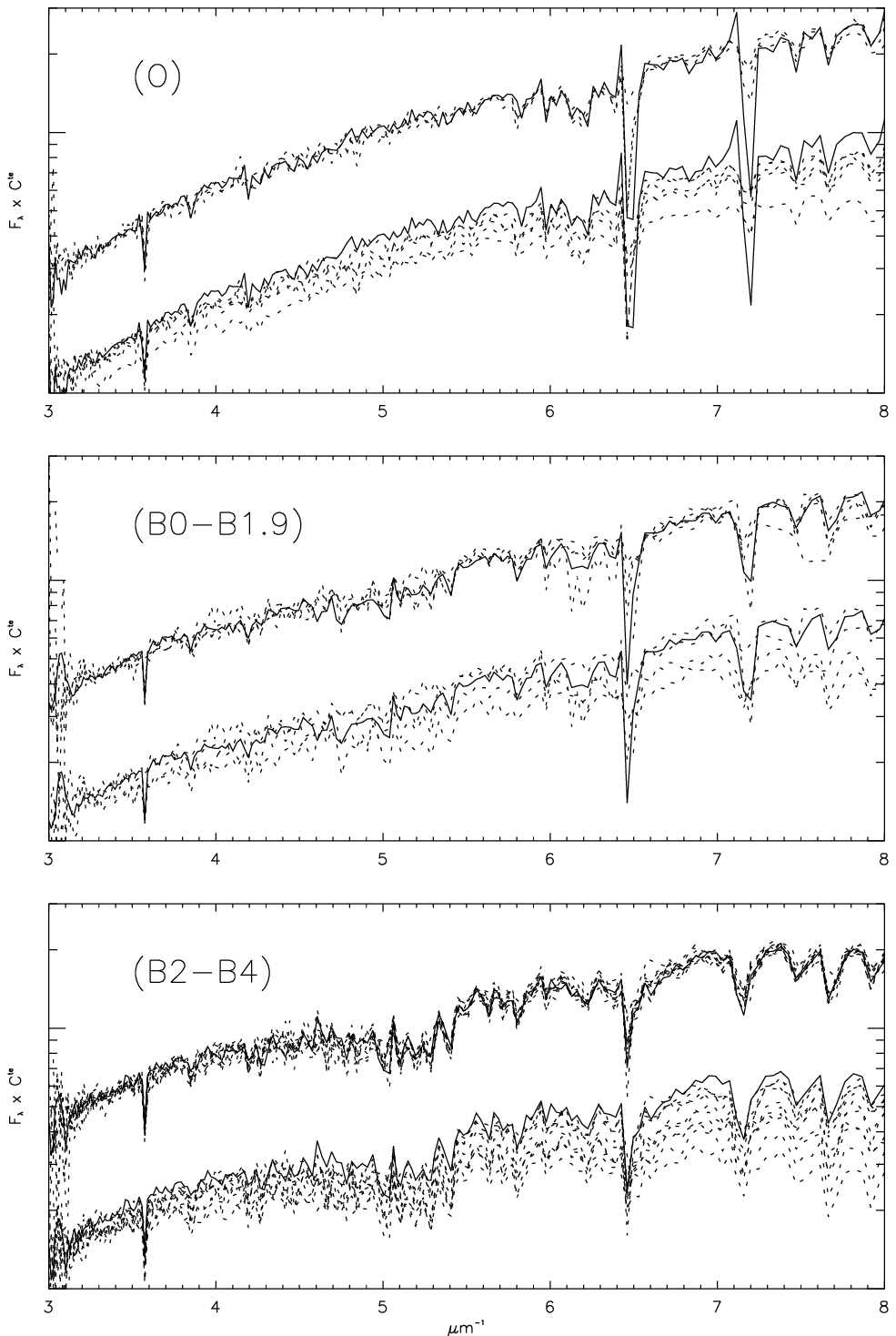}} 
\caption{Slightly reddened Magellanic stars ($B-V<0$, Sect.~\ref{bvn}) of Table~\ref{tbl:vsr}, classified by spectral type.
On each plot (Y-logarithmic axis),  bottom spectra are the UV spectra re-scaled to a common value in the near-UV,
top spectra are the same spectra, corrected for their difference of slope.
Solid lines are for Sk-67 168 (O), Sk-68 41 (B0-B1.9), Sk-65 15 (B2-B4).} 
\label{fig:type}
\end{figure}
\begin{figure}
\resizebox{1.\columnwidth}{!}{\includegraphics{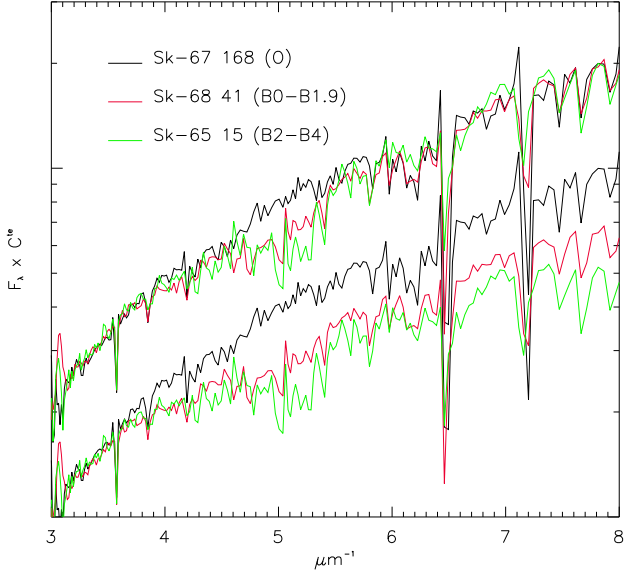}} 
\caption{
Bottom spectra are IUE spectra of Sk-67~168 (O type, black line), Sk-68~41 (B0-B1.9, red line), Sk-65~15 (B2-B4, green line), scaled to the same arbitrary value in the 
near-UV.
Top, spectra of Sk-68~41 and of Sk-65~15 have been multiplied 
respectively by $e^{0.08/\lambda}$ and  by $e^{0.14/\lambda}$, to 
match the slope of the spectrum of Sk-67~168. Y-logarithmic axis.}
\label{fig:typec}
\end{figure}

\begin{table*}
\caption[]{Slightly reddened stars ($B-V<0)$}		
       \[
    \begin{tabular}{c|c|cc|cc}
\hline
star  &  $E(B-V)^{(1)}$ & $\Delta_{vis}^{(2)}¥$&$\Delta_{uv}^{(3)}$
 & $\Delta_{vis}^{(4)}$&$\Delta_{uv}^{(5)}$
 \\
\hline
\multicolumn{6}{c}{O stars}\\ 
\hline
Sk-67 168 &0.12 & 0 & 0 & 0 & 0
\\
AzV70 &0.12 & 0.032 &  0.025&&
\\
Sk-65 63 & 0.11 & 0.02 & 0.04&&
\\
Sk-66 169 &0.14 & 0.08 & 0.05 &&
\\
AzV289 &0.15 & 0.10 & 0.11&&
\\
\hline
\multicolumn{6}{c}{B0-B1.9}\\ 
\hline
AzV462& 0.04  & 0 & 0 &&
\\
Sk-68 41 & 0.07& 0.03 & 0.01  & 0.06 & 0.08
\\
Sk-67 5 & 0.13& 0.07 &0.05&&
\\
AzV380 & 0.12  & 0.032  & 0.056&&
\\
AzV404 &0.05 & 0.056 & 0.074&&
\\
\hline
\multicolumn{6}{c}{B2-B4}\\ 
\hline
Sk-65 15 &  0.07 & 0  & 0  & 0.12 & 0.14
\\
Sk-66 35 &  0.10 & 0.06 & 0.00 &&
\\
Sk-67 256 &  0.09 & 0.06 &  0.01&&
\\
Sk-67 100 &0.1 & 0.04 & 0.04&&
\\
Sk-70 120  & 0.12 & 0.08 & 0.03&&
\\
Sk-66 106 &  0.08 & 0.06 & 0.04 &&
\\
Sk-67 36 &  0.08& 0.06&0.06&&
\\
Sk-66 118 & 0.10& 0.12 &0.08&&
\\
Sk-67 228 & 0.12 &0.12 & 0.09&&
\\
Sk-68 40 &  0.08 &0.08&0.12 &&
\\
Sk-67 78 &  0.09 & 0.14&  0.15&&
\\
\hline
\end{tabular} 
 \]
\begin{list}{}{}
\item[$(1)$] Rough estimate of the reddening from Tables~\ref{tbl:smc} and \ref{tbl:lmc}.
\item[$(2)$] Twice the difference of B-V between the star and the 
reference star of same type. Corresponds to the slope difference of 
the two stars spectra in the visible.
\item[$(3)$] Difference of UV slopes between the star and the reference 
star of same type. This is the exponent found in Figure~\ref{fig:type} 
for the UV spectra.
\item[$(4)$]Twice the difference of B-V between the star and Sk-67 168.
\item[$(5)$] Difference of UV slopes between the star and Sk-67 168.
From Figure~\ref{fig:typec}.
\end{list}
\label{tbl:vsr}
\end{table*}
Figure~\ref{fig:type} demonstrates that the UV 
spectra of these slightly reddened stars (with reddenings 
of order $E(B-V)\sim 0.1$~mag., Table~\ref{tbl:vsr}), within each spectral type (O, B0-B1.9, B2-B4, Sect.~\ref{uv}), differ 
exactly by a change of slope.
The bottom spectra on each plot are the UV spectra of the stars rescaled to the same value in the near-UV.
The top spectra show the good superimposition (B2-B4 stars in particular, once the slope differences are corrected, have identical continuum and  spectral lines strengths) of the spectra, once they have been corrected for their slope differences.
Interstellar extinction should be the main cause of change of slope within 
each of the three spectral types in Figure~\ref{fig:type}.
Modification of the UV continuum between spectral types is also an exponential of $1/\lambda$ (Figure~\ref{fig:typec}).

The perfect superimposition of the UV spectra within each class, once 
they have been corrected for their slope differences, 
implies (Sect.~\ref{elin}) that the interstellar extinction curves in the UV are, for these 
directions, linear in $1/\lambda$.
It will also be deduced that Magellanic and Galactic UV extinction laws in these 
directions (unless they can compensate one another to  give the observed linear extinction law)   both behave  as $1/\lambda$ and are not discernable.

There is also excellent agreement between  slope differences in the visible, $\Delta_{vis}=2\Delta(B-V)$ (column~3, Table~\ref{tbl:vsr}), and in the UV ($\Delta_{uv}$, column~4, with an accuracy of $\sim 10\%$) within each class of stars (and also between the different classes, last columns of the Table).
Therefore, the extinction law in these directions is linear across the whole visible and UV wavelength range. 
\section{Stars with no $2200\,\rm\AA$ bump, and $B-V>0$} \label{nb}
\begin{figure}
\resizebox{1.\columnwidth}{!}{\includegraphics{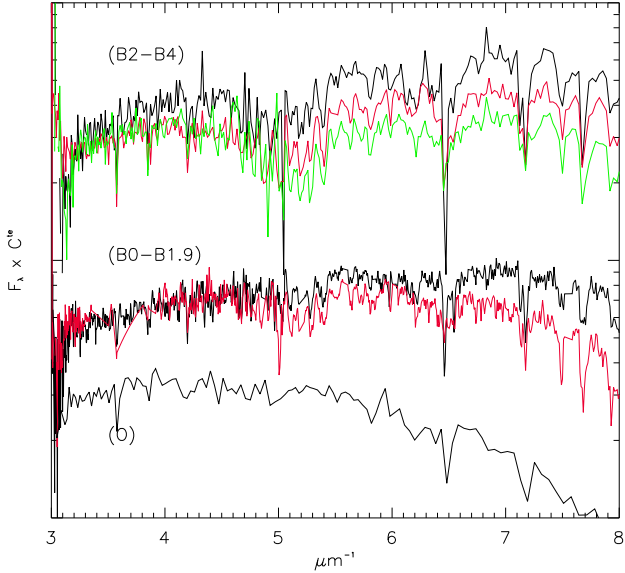}} 
\caption{UV spectra of reddened stars, with $B-V>0$ and no 
$2200\,\rm\AA$ bump (Sect.~\ref{nb}).
AzV398 is the only O star.
B0-B1.9 stars are AzV18 (in red) and AzV214 (in black).
B2-B4 are Sk-69~256 (black line), Sk-69~228 (red line), Sk-70~116 (green line).
 Y-logarithmic axis.} 
\label{fig:nb1}
\end{figure}

A few LMC/SMC stars have a positive $B-V$ (which 
for these O-B stars is proof of a non negligible reddening) and,
contrary to common Galactic stars with similar reddening, have no $2200\,\rm\AA$ bump.
These are AzV398 O), AzV18 and AzV214 (B0-B1.9), in the SMC; Sk-69~228, Sk-69~256, 
and Sk-70~116 (B2-B4), in the LMC (Figure~\ref{fig:nb1}). 
$E(B-V)$ ranges from 0.2 to 0.4~mag., significantly larger than for  the slightly reddenned stars of Sect.~\ref{bvn}.

\begin{figure}[t]
\resizebox{1.\columnwidth}{!}{\includegraphics{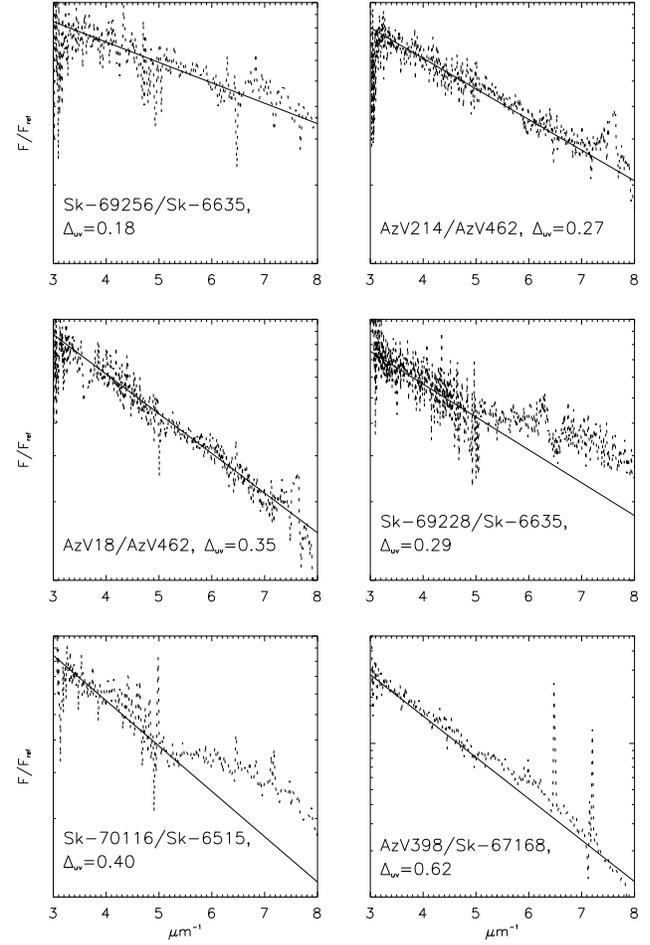}} 
\caption{Reduced spectra for the stars of Figure~\ref{fig:nb1},  Y-logarithmic axis.
The straight line on each plot, with slope $\Delta_{uv}$, prolongs the visible extinction.
The y-axis spans a factor of 10 for all but AzV398 plots, a factor of 
15 for AzV398's.
}
\label{fig:nb2}
\end{figure}

Normalisation of the UV spectra of the stars  (Figure~\ref{fig:nb2}) by the low 
reddened ones of  Sect.~\ref{bvn}, of same type, gives in the near-UV an exponential 
of $1/\lambda$ (the exponent $\Delta_{uv}$ is written on each plot).

For three of the stars (Sk-69~256, AzV214, AzV18) this 
exponential can be  prolonged  to the far-UV.
The increase of reddening from  Sk-69~256 to AzV18 
is clearly observed through the progressive change of slope from one reduced spectrum to the other  (Figure~\ref{fig:nb2}).
There is excellent agreement between $\Delta_{uv}$ and $\Delta_{vis}$: the extinction law in these directions is linear from the visible to the far-UV, and, as in Sect.~\ref{bvn}, Galactic and Magellanic extinctions along these lines of sight can not be distinguished.

For AzV398, Sk-69~228 and Sk-70~116, $\Delta_{vis}$ is still a good indicator of 
the difference of near-UV slope with stars of lesser reddening, but in the far-UV the SEDs ratios are in excess compared to the visible slopes.
The exact wavelength from which the curves depart from linearity is not clear (especially for Sk-70~116) owing to the overlap between the bump and the AL/Fe absorption regions.
\section{Reddened stars with a $2200\,\rm\AA$ bump} \label{bump}
Reddened stars with a $2200\,\rm\AA$ bump are all but one (AzV398) in the LMC.
They will be distinguished according to their spectral type  and to the size of the bump (weak or large).
\subsection{Weak bumps} \label{sb}
\begin{figure}
\resizebox{1.\columnwidth}{!}{\includegraphics{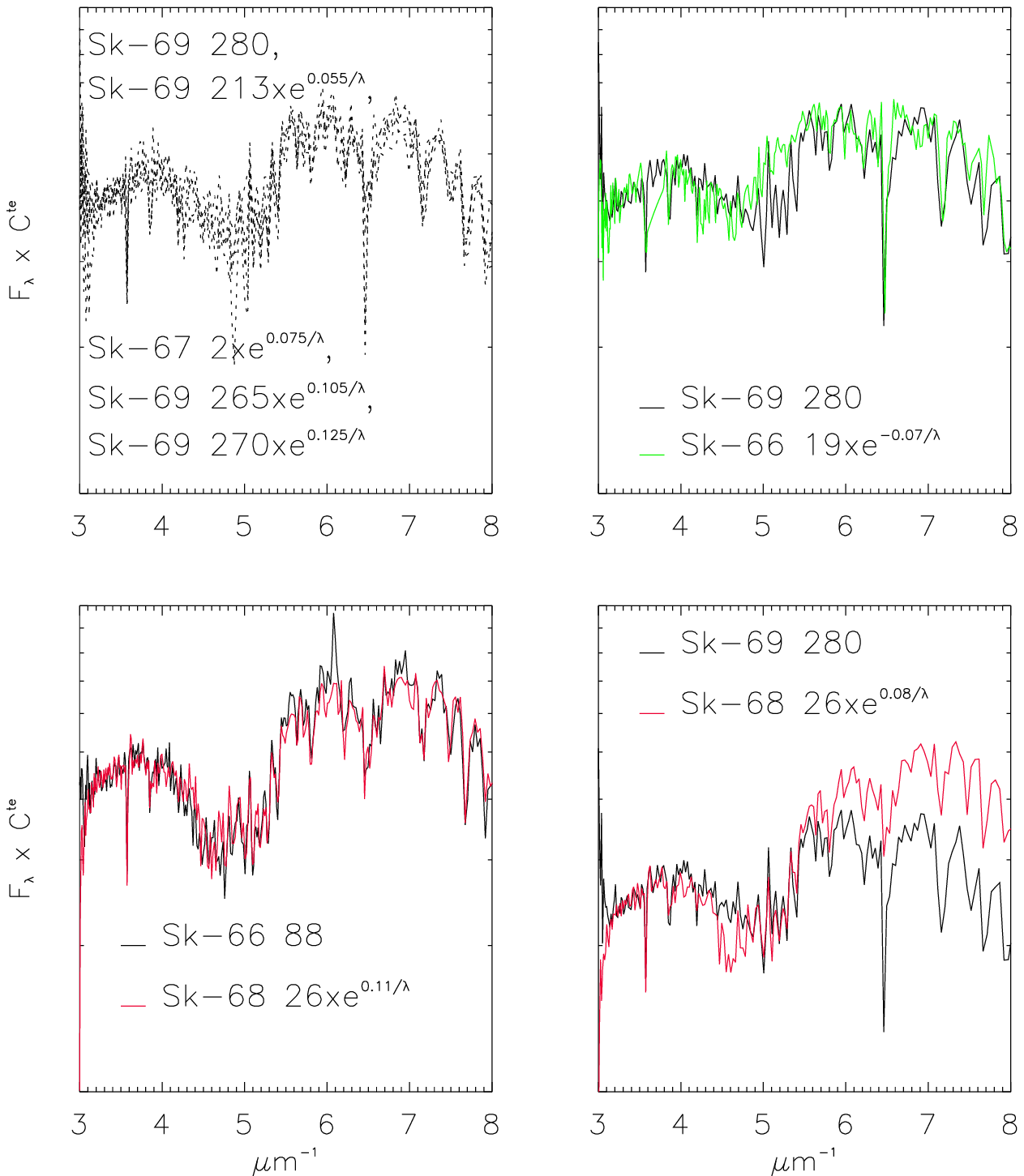}}
\caption{Stars with a weak bump (Sect.~\ref{sb}). Y-logarithmic axis.
} 
\label{fig:mb}
\end{figure}
Among  the twelve stars with a weak bump (quoted by a 'W', last column in Tables~\ref{tbl:smc} and \ref{tbl:lmc}), nine [Sk-68~155 (O), Sk-66~19,  Sk-68~129, Sk-68~140 (B0-B1.9), Sk-69~213, Sk-69~280, Sk-67~2, Sk-69~265, Sk-69~270 (B2-B4)] have UV spectra which differ exactly by a change of slope (an exponential of $1/\lambda$, Figure~\ref{fig:mb}, top plots). 
The correspondence between the change of slope in the visible and in the UV between the stars  (last column of Table~\ref{tbl:wb}) is excellent for all stars but Sk-66~19.
Along all these directions (except Sk-66~19) reduced spectra will differ by an exponential of $1/\lambda$, and the extinction laws by a linear term, over all the visible-UV wavelength range.

The UV spectra of  Sk-66~19 and of Sk-69~280 superimpose perfectly with a slight change of slope (right top plot of Figure~\ref{fig:mb}).
A break of this change of slope between the visible and the near-UV can not be excluded although it would represent an exception among the stars considered so far.
On the other hand, a slightly larger error than expected in Sk-66~19's photometry, or the possibility that $B-V$ does not properly represent its slope would explain the discrepancy.
A comparison of the visible SEDs of the stars would best fix this problem.

B2-B4 stars Sk-66~88, Sk-68~26, AzV456, lie apart.
Their visible/UV spectra differ one from another  by a change of slope, which is equal (within the error-margins) in the visible and in the UV (Table~\ref{tbl:wb}).
Their UV spectra do not match the preceding ones, whichever correction of the slope is applied:
 if, for instance, the UV spectrum of Sk-66~88, corrected for the difference of slope given by the visible photometry, is compared to the spectrum of Sk-68~129 (Figure~\ref{fig:mb}, bottom right plot), large differences appear in the far-UV part of the spectra (the bump may also be smaller for Sk-66~88 than it is for Sk-68~129). 
The implication, already apparent  in Sect.~\ref{nb}, is that visible extinction alone does not suffice to fix the extinction in the UV. 
One additional parameter, at least, is necessary.
However linear extinction laws still play a role in these directions of higher extinction since the visible-UV spectra of many of them deduce one from the other by an exponential of $1/\lambda$.
\begin{table}
\caption[]{Stars with a weak bump ($0.15<B-V<0.4$)}		
       \[
    \begin{tabular}{ccccc}
\hline
star  & type & $\Delta_{vis}^{(1)}$ & $\Delta_{uv}^{(2)}$ & {$(\Delta_{vis}-\Delta_{uv})$}
\\
\hline
Sk-68 155  & O & -0.06 & -0.05&-0.01
\\
Sk-66 19 & B0-B1.9 & 0.06  & -0.07 &  0.13 
\\
Sk-68 129 &B0-B1.9 & -0.12  & -0.15 &  -0.03
\\
Sk-68 140 & B0-B1.9 &0.00 & 0.06   & -0.06  
\\
Sk-69 280  & B2-B4 & 0  & 0  & 0 
\\
Sk-69 213 & B2-B4  & 0.02  & 0.05  & -0.03 
\\
Sk-67 2  & B2-B4 & 0.02  &  0.07 & -0.05 
\\
Sk-69 265  & B2-B4  & 0.06  & 0.10  & -0.04
\\
Sk-69 270  & B2-B4  &  0.10 &  0.12 & -0.02 
\\
\hline
Sk-68 26  &  B2-B4 & 0  &  0 & 0
\\
Sk-66 88  & B2-B4  &  0.14 & 0.11  & 0.04  
\\
AzV 456 & B0-B1.9 & -0.06  & -0.07  & -0.01 
\\
\hline
\end{tabular} 
 \]
\begin{list}{}{}
\item[$(1)$] Twice the difference of B-V between the star and the star chosen as reference (the star with $\Delta_{vis}=\Delta_{uv}=0$).
\item[$(2)$] Difference of UV slopes between the star and the reference one. 
\end{list}
\label{tbl:wb}
\end{table}
\subsection{Large bumps} \label{lb}
\begin{figure}[t]
\resizebox{1.\columnwidth}{!}{\includegraphics{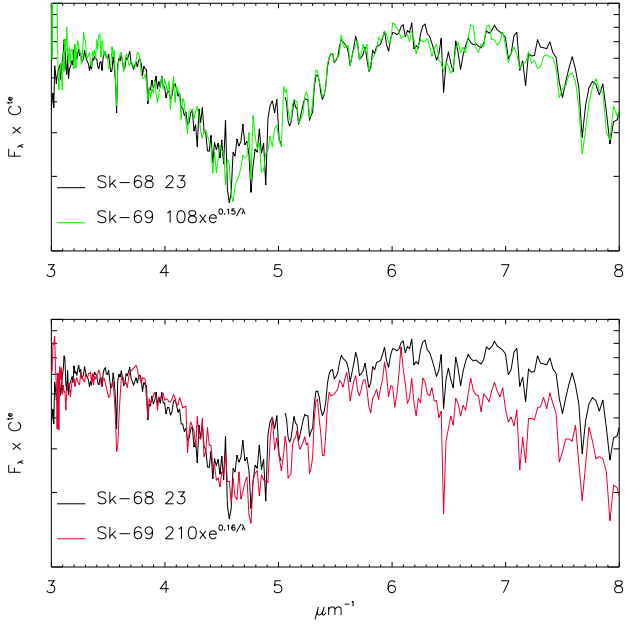}} 
\caption{Stars with a large bump (Sect.~\ref{lb}). Y-logarithmic axis.
} 
\label{fig:gb}
\end{figure}
Most pronounced bumps are observed in the spectra of B2-B4 stars with relatively high reddenings ($E(B-V)\sim 0.5$) Sk-68~23, Sk-69~108, and Sk-69~210.

Sk-69~108 and Sk-68~23 have a slightly larger bump than Sk-69~210.
Their UV spectra differ by an exponential 
$e^{0.15/\lambda}$, the exponent being close to the change of 
visible slope $\Delta_{vis}
 \sim 0.10$ (Figure~\ref{fig:gb}, top).

The bottom plot of Figure~\ref{fig:gb} proves that the UV spectra of Sk-68~23 and Sk-69~210 superimpose well in the near-UV, after the slope of the latter has been corrected for the difference of visible reddening, but the spectra do not match in the far-UV.
There is no straightforward relationship between the variation of the
extinction curve in the visible and in the UV between the directions of Sk-69~210 on the one hand, and Sk-68~23 and Sk-69~108 on the other.
\section{Fit of LMC/SMC extinction curves} \label{ec}
\subsection{The CCM parametrization} \label{ccm}
\begin{figure}[t]
\resizebox{1.\columnwidth}{!}{\includegraphics{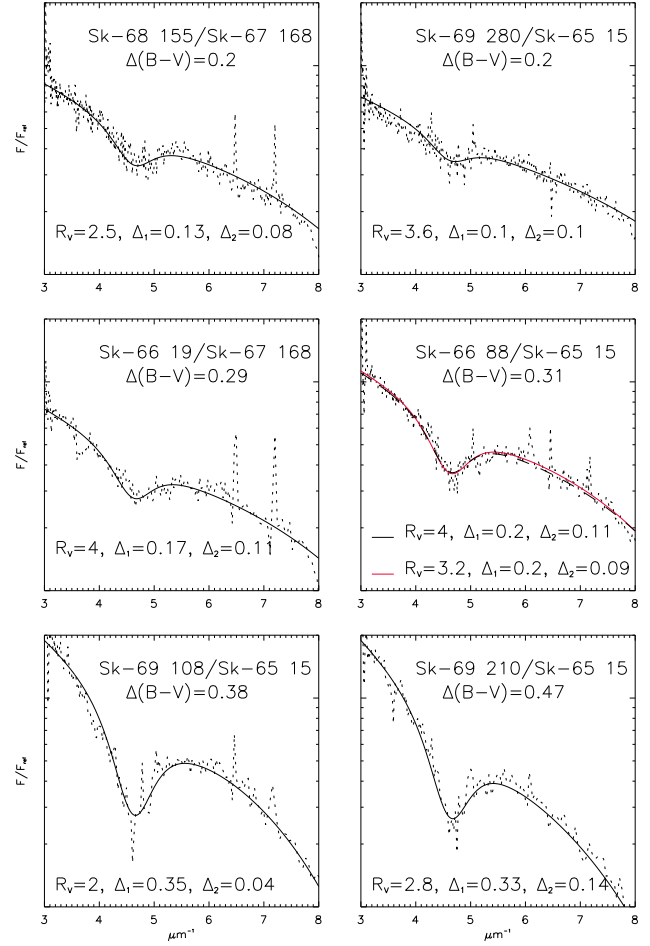}} 
\caption{Fit (Eq.~\ref{eq:ccmg}) of the  reduced spectra of a representative set of Magellanic stars with a bump (Sect.~\ref{fit}).
 Y-logarithmic axis.
} 
\label{fig:mcfit}
\end{figure}
Figures~1 and 2 in CCM demonstrate that the different parts of Galactic extinction curves vary simultaneously, in a systematic way, depending on $E(B-V)$ and one additional parameter only: at all wavelengths and for all directions, there is a linear relationship between $A_\lambda/E(B-V)$ and $R_V^{-1}=A_V/E(B-V)$.
Since $E(B-V)$ can be simplified between these two expressions, the CCM finding means that $A_\lambda$ is not determined by $E(B-V)$ alone, but is a linear combination  of $E(B-V)$ and $A_V$.
In other words, the exact shape of the extinction curve of an average Galactic star is fixed by the visible reddening, and the knowledge of one point (here ($\lambda_V, A_V$)) in the $(1/\lambda,\, A_\lambda)$ plane.
The mathematical transcription of the CCM finding is that there are two functions $a(x)$ and $b(x)$ which satisfy 
\begin{eqnarray}
\frac{A_{\lambda}}{A_V}&=&a(x)+b(x)R_V^{-1} \nonumber \\
    \label{eq:ccm0}
\end{eqnarray}
(eq.~1 in CCM, $x=1/\lambda$) or
\begin{eqnarray}
A_{\lambda}&=&a(x)A_V+b(x)E(B-V) \nonumber \\
&=&E(B-V)(a(x)R_V+b(x))
    \label{eq:ccmf}
\end{eqnarray}
In the UV $a(x)$ and $b(x)$ are given by eqs.~4a and 4b in CCM
\begin{eqnarray}
a(x)&=&1,752-0.316-\frac{0.104}{(x-4.67)^2+0.341}+F_a(x) \nonumber \\
b(x)&=&-3.090+1.825x+\frac{1.206}{(x-4.62)^2+0.263}+F_b(x) \nonumber 
    \label{eq:ab}
\end{eqnarray}
with
\begin{eqnarray}
F_a(x)&=&-0.0447(x-5.9)^2-0.009779(x-5.9)^3  \nonumber \\
&&(8\geq x \geq 5.9)   \nonumber  \\
F_b(x) &=& 0.2130(x-5.9)^2 + 0.1207(x-5.9)^3  \nonumber \\
&&(8 \geq x  \geq 5.9)  \nonumber \\
F_a(x)&=&F_b(x)=0  \nonumber \\
&&   (x <5.9) \nonumber 
    \label{eq:f}
\end{eqnarray}
Eq.~\ref{eq:ccmf} is equivalent to
\begin{eqnarray}
\frac{F_{\star}}{F_{0}}&=&Ce^{-0.92E(B-V)(b+aR_V)} \nonumber \\
&=&Ce^{-0.92E(B-V)b(x)}e^{-0.92E(B-V)R_Va(x)}
    \label{eq:ccmr}
\end{eqnarray}
($F_\star$ and $F_{0}$ are the reddened and the reference stars spectra, $C$ a constant).

Two remarks should be made.
One is that in practice, taking the $a(x)$ and $b(x)$ functions given in CCM (and reproduced above),  the information on the $2200\,\rm\AA$ bump feature is nearly all contained in the first exponential of Eq.~\ref{eq:ccmr}, which fixes the size of the bump, and an average slope for the extinction curve.
Therefore, in the CCM framework, the bump depends mainly on $E(B-V)$ (almost not on $R_V$), an hypothesis only partially confirmed by observation \citep{savage75}.
The second exponential, thus the value of $R_V$, allows some variation of the UV curve's slope around the mean slope. 

The second point that we must draw attention to is that linear extinction laws  in the UV, as they were encountered in previous sections, are not accounted for by the CCM fit: it would require the exponent involving $b(x)$ to be null (no bump), and thus: $E(B-V)=0$.
Therefore most of SMC extinction curves, and Galactic sight-lines of very low column densities \citep{z01} will not be fitted by the CCM relationship.
\subsection{Fit of extinction curves} \label{fit}
\begin{figure*}
\resizebox{1.\textwidth}{!}{\includegraphics{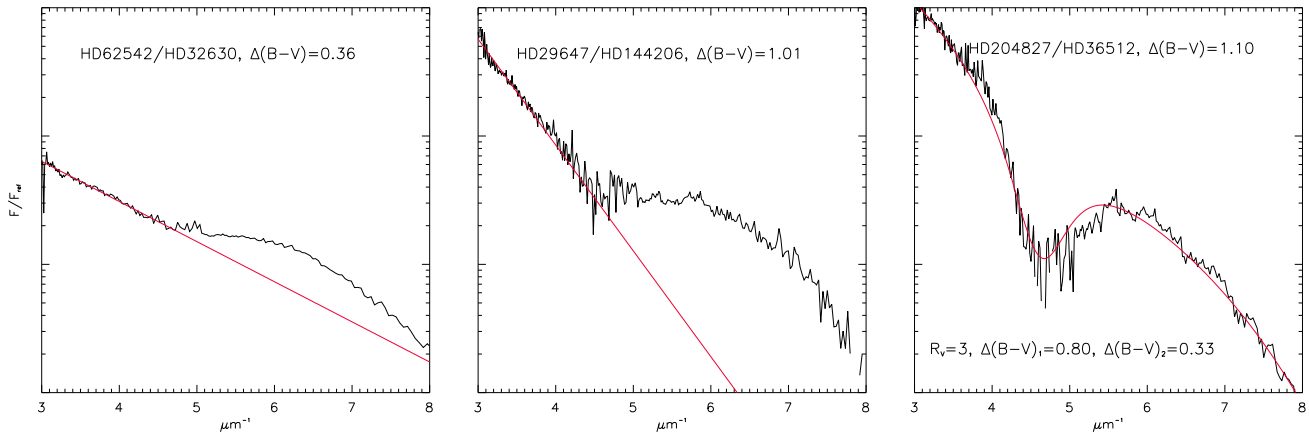}} 
\caption{Reduced spectra of Milky Way stars which do not follow the CCM relationship \citep{valencic04} (Sect.~\ref{fitr}).
The straight line on the two first plots is the UV prolongation of the visible extinction. HD204827 is fitted by Equation~\ref{eq:ccmg}.
 Y-logarithmic axis.} 
\label{fig:mwfit}
\end{figure*}
\citet{g03} finds that four directions only, out of the forty one of the sample, Sk-69~280, Sk-66~19  (weak bump, Sect.~\ref{sb}), Sk-68~23, Sk-69~108   (large bump, Sect.~\ref{lb}) follow a CCM relationship (with respective $R_V$ of 3.12, 3.44, 3.55, 3.15).
It follows from Sect.~\ref{bump} that all stars with a bump whose spectra differ from these four ones by a change of slope only (an exponential of $1/\lambda$ over all the visible wavelength range) will have their reduced spectrum fitted by the product of a CCM fit (Eq.~\ref{eq:ccmr}) and an exponential of $1/\lambda$: the extinction law in these directions (that is all directions with a bump except these of Sk-68~26, Sk-66~88, AzV456, Sk-69~210 and maybe Sk-66~19) is the sum of a CCM and a linear  extinction laws.
For these directions, it is possible to find $\Delta_1$ and $\Delta_2$ for which
\begin{eqnarray}
\frac{F_{\star}}{F_{0}}&=&Ce^{-0.92\Delta_1(b+aR_V)}e^{-2\Delta_2/\lambda} \nonumber \\
&=&Ce^{-0.92\Delta_1b(x)}e^{-0.92\Delta_1R_Va(x)}e^{-2\Delta_2/\lambda},
    \label{eq:ccmg}
\end{eqnarray}
with $\Delta_1+\Delta_2\sim \Delta(B-V)$.

From Figure~\ref{fig:mcfit} it will be concluded  that this fit also applies to Sk-66~19, Sk-68~26 (thus also to Sk-66~88, AzV456, Sect.~\ref{sb}) and to Sk-69~210.
Since it holds for directions with a linear extinction, it finally applies to all forty one stars of the sample, with the exception of  Sk-69~228, Sk-70~116, and AzV398 which do not exhibit the bump feature and do not have a linear extinction.

$\Delta_1$ in Eq.~\ref{eq:ccmg} is unambiguously fixed by the strength of the $2200\,\rm\AA$ bump, $\Delta_2$ by the difference between $\Delta(B-V)$ and $\Delta_1$.
$R_V$  then adjusts the UV slope of the fit, as in the CCM framework (Sect.~\ref{ccm}).
\begin{figure*}[t]
\resizebox{1.\textwidth}{!}{\includegraphics{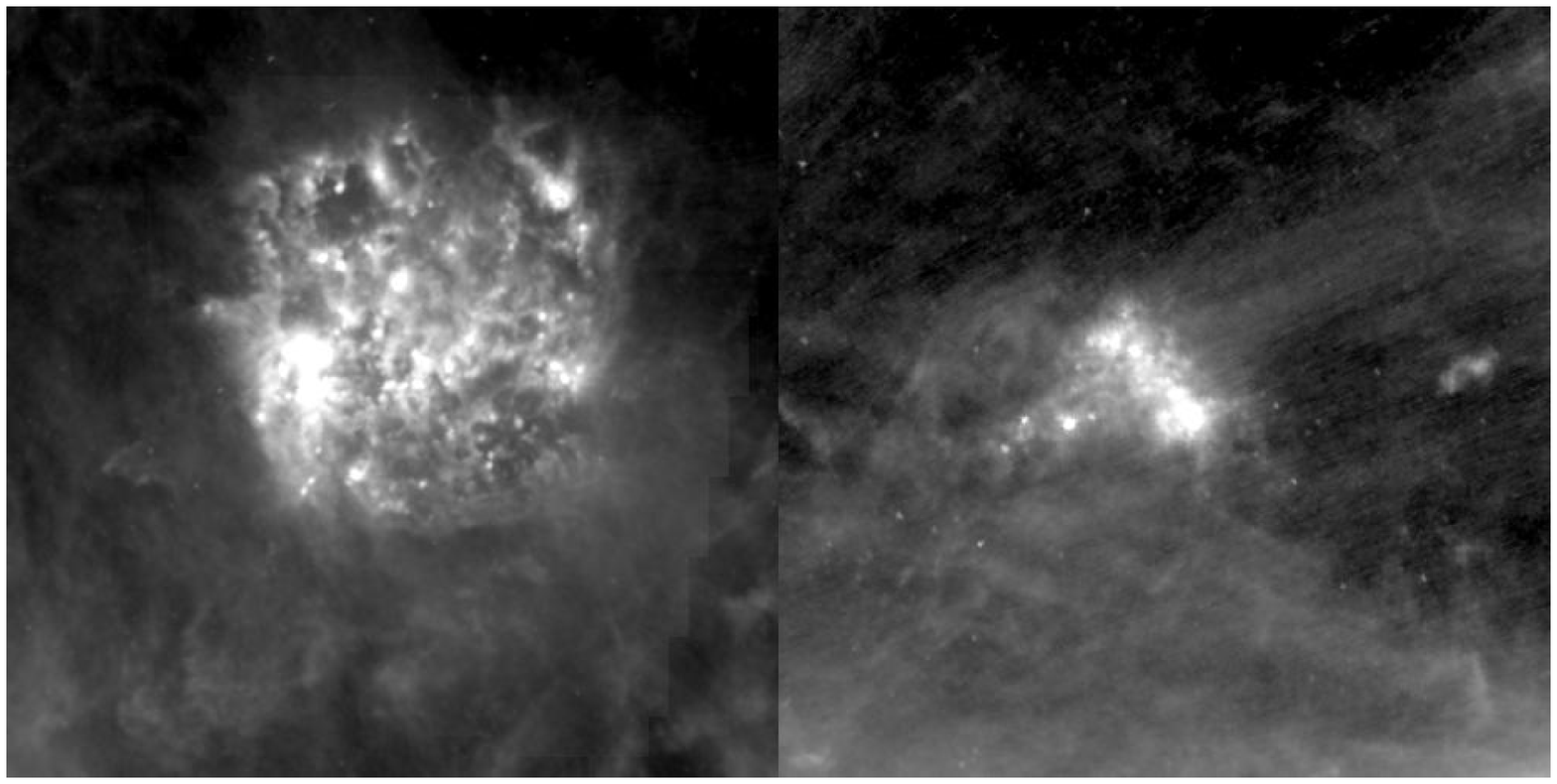}} 
\caption{IRAS $100\,\rm\mu m$ images (gnomonic J2000 coordinates, 1~pix=90", each field is $12.5^{\circ}\times 12.5^{\circ}$ wide) of the LMC (left) and the SMC (right) (from \emph{Sky-View}). The field, larger than the extent of the galaxies, includes the Galactic cirrus on each line of sight. Cutoffs are 1.5~MJy/sr and 100~MJy/sr for the LMC, 0.5 and 30~MJy/sr for the SMC. Logarithmic scale.
} 
\label{fig:mc100}
\end{figure*}

The expression of a normalized extinction curve can easily be  
retrieved from  Eq.~\ref{eq:ccmg}:
\begin{eqnarray}
\frac{A_{\lambda}}{E(B-V)}&=& \frac{\Delta_1}{E(B-V)}(a(x)R_V+b(x)) \nonumber \\
&&+\, 0.53(1- \frac{\Delta_1}{E(B-V)})x,
    \label{eq:ccmgm}
\end{eqnarray}
$\Delta_1$ being unequivocally determined by the size of the  $2200\,\rm\AA$ bump.
\subsection{Remarks on Eqs.~\ref{eq:ccmg} and \ref{eq:ccmgm}} \label{fitr}
Eq.~\ref{eq:ccmg}  (or \ref{eq:ccmgm}) separates, along each sight-line, a CCM component from a linear component.
As in CCM, the bump size is associated with a value ($\Delta_1$) of the reddening  although, in this new formulation, it may be only part of $E(B-V)$.
The remaining part  ($\Delta_2=E(B-V)-\Delta_1$) of $E(B-V)$ follows a linear extinction law.
The correspondence between $E(B-V)$ and bump size, implicit in CCM, is thus a boarder case ($\Delta_1\rightarrow E(B-V)$, $\Delta_2\rightarrow 0)$, rather than the generality.

Firstly, this allows one to account  for linear extinction curves.
Secondly, it allows a fit of most Galactic and Magellanic extinction curves by a single, common mathematical expression.
For a given bump size the number of free parameters of a normalized extinction curve remains unchanged (=~1).

There is, for this fit as well as for CCM's, a large uncertainty on the value of $R_V$ due to the uncertainty on $\Delta (B-V)$:
middle right plot of Figure~\ref{fig:mcfit} gives two acceptable and nearly undistinguishable fits of the UV spectrum of Sk-66~88 which differ by $\Delta(B-V)= 0.02$ while the values found for $R_V$ vary from 3.2 to 4; any small uncertainty on $\Delta(B-V)$  leads to a large variation of $R_V$.
The sensitivity of the $R_V$ parameter to the reference star is also indicated by a comparison of the values given in \citet{g03} to the ones found here: for   Sk-66~19, for instance, with  Sk-66~169 as comparison, \citet{g03} finds $R_V=3.44\pm 0.21$ while the present decomposition, with Sk-67~168 as reference, gives $R_V\sim 4$.
Noteworthy, \citet{bouchet85} finds, from a multi-wavelength analysis, $R_V=2.7\pm 0.2$ in the SMC.

Eq.~\ref{eq:ccmg} can be used to re-examine particular Galactic extinction curves.
 Only four directions out of the 417  studied by \citet{valencic04} do not follow a CCM relationship.
The reduced spectra of three [HD62542 ($E(B-V)=0.38$), HD204827 ($E(B-V)=0.80$), HD29647 ($E(B-V)=0.91$] of these directions (the fourth, HD210121, is not considered here because its spectral type and $B-V$ color are poorly known) are plotted on Figure~\ref{fig:mwfit}.
HD204827 has a bump and is well fitted using Eq.~\ref{eq:ccmg}  with $R_V\sim 3$.
The two other directions, HD62542 and HD29647, have no or little bump, and their near-UV reduced spectra prolong the visible extinction law, the far-UV part being in excess.
As  Sk-69~228, Sk-70~116, AzV398 they can not be fitted by  Eq.~\ref{eq:ccmg}.
\section{On the separation of Galactic and Magellanic extinctions} \label{esep}
Sects.~\ref{bvn}, \ref{nb}, and \ref{fit} demonstrate that it is not possible to separate analytically Galactic and Magellanic extinctions. 
An attempt to bypass this difficulty is made in \citet{misselt99}: reddened stars in the LMC are associated to reference stars with a similar Galactic reddening, estimated from the  \citet{ostreicher95} $10'$ resolution Galactic map.
According to this paper the resulting extinction curves isolate the Magellanic extinction. 

\begin{figure}[t]
\resizebox{\columnwidth}{!}{\includegraphics{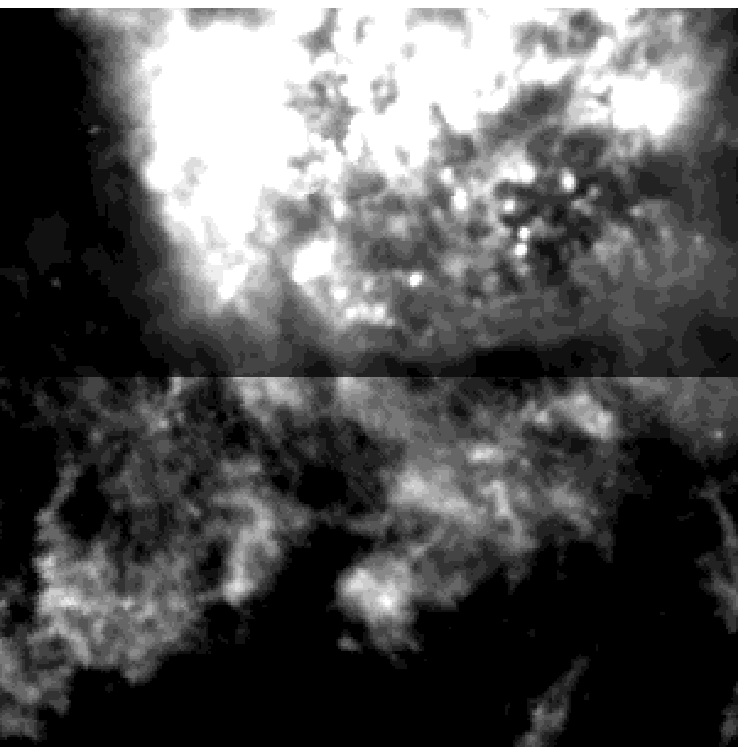}} 
\caption{ Bottom two thirds of LMC image (Figure~\ref{fig:mc100}), with different cutoffs (3 and 10~MJy/sr for the lowest part, 3 and 30~MJy/sr for the upper one).
There is continuity of the infrared structures in and out of the LMC.
} 
\label{fig:lmc100}
\end{figure}

Both the LMC and the SMC are observed behind Galactic cirrus.
The highlighted infrared (filamentary) structures (see Figure~\ref{fig:mc100}, retrieved from \emph{Sky-View} at http://skys.gsfc.nasa.gov) observed in the direction of the Clouds belong to the Galaxy: they reproduce the structure of all Galactic cirrus observed by IRAS; there is continuity between the filaments in and outside the Magellanic Clouds (Figure~\ref{fig:lmc100}); and it would be remarkable that such structures, which would represent enormous masses of gas at the distance of the Magellanic Clouds, can mimic so well Galactic cirrus (unless one supposes self-similarity of cirrus up to very large scales).
The large infrared surface brightness at their position results from  the strong infrared gradients due to dust surrounding the Magellanic giant stars, over-which is superimposed the much lower Galactic HLC emission.
\begin{figure}[t]
\resizebox{\columnwidth}{!}{\includegraphics{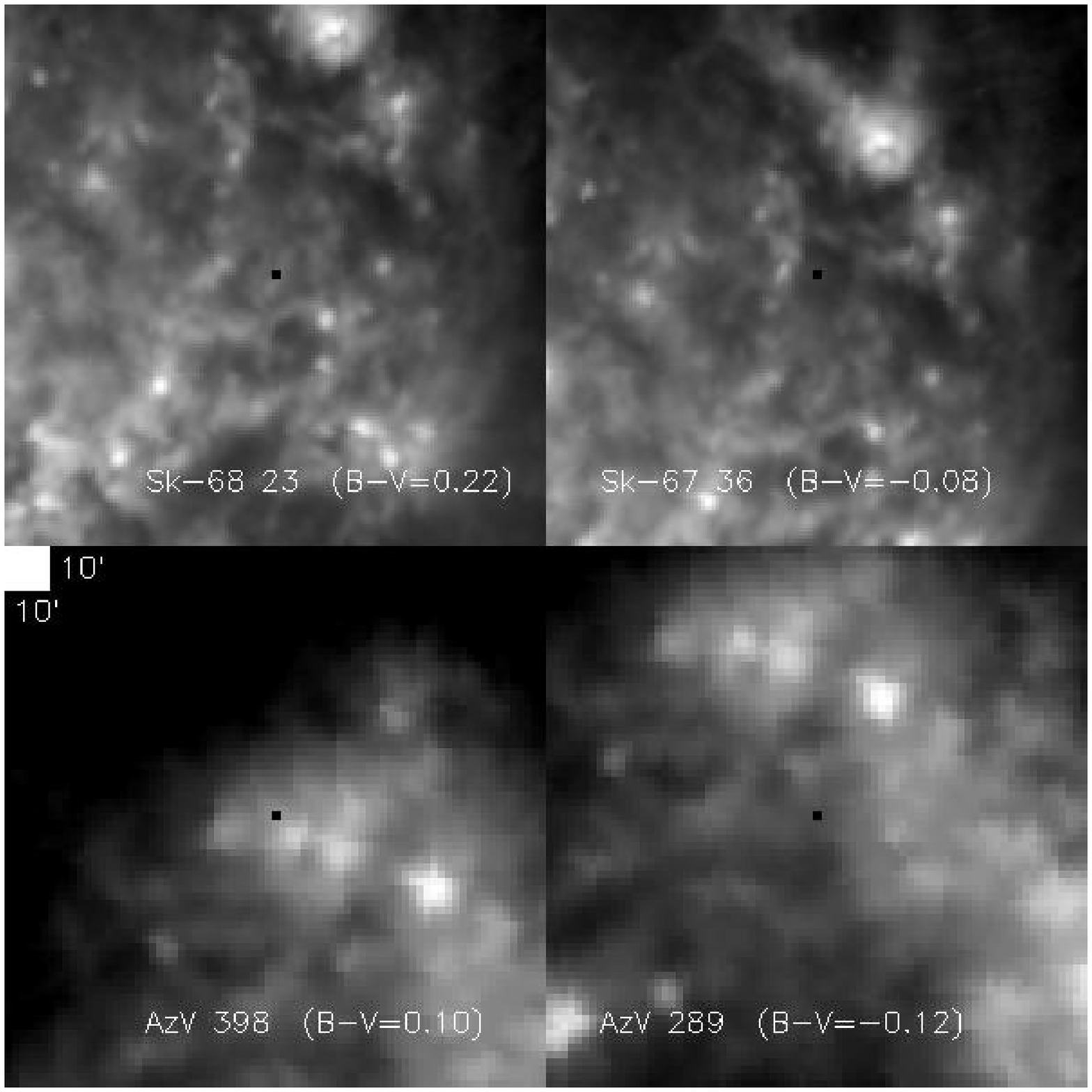}} 
\caption{Four times magnified IRAS  $100\,\rm\mu m$ images (from \emph{Sky-View}) centered on reddened  Sk-68~23 (top left) and AzV~398 (bottom left), and their comparison stars in \citet{g03}, Sk-67~36 (top right) and AzV~289 (bottom right).
Each image is $1.87^{\circ}\times 1.87^{\circ}$ wide, cutoffs are identical for reddened and comparison star images.
The beam of \citet{ostreicher95} extinction map  is given by the bright square top left of the left bottom image.
} 
\label{fig:100muc}
\end{figure}

It will be deduced that the Galactic extinction map in these directions is structured down to very small scales: at least the one arcminute probed by IRAS images, and most probably to sub-arcseconds as it can be observed on visible images of Galactic cirrus \citep{z99}.
The relative positions of the Magellanic stars and of Galactic structures on IRAS maps can serve as an indicator of the different amounts of Galactic dust along the stars' directions.
Figure~\ref{fig:100muc} compares the $100\,\rm\mu m$ emission around reddened
Sk-68~23 (top left) and AzV~398 (bottom left), and the comparison stars adopted in \citet{g03}, respectively Sk-67~36 (top right) and AzV~289 (bottom right).
In both cases the reddened stars are situated behind local enhancements of Galactic interstellar matter, while the comparison stars are seen through 'holes' (presumably) of much lower column densities.
The $10'$ resolution (see Figure~\ref{fig:100muc}) of  the \citet{ostreicher95} Galactic extinction map in the direction of the LMC lacks the precision needed to resolve the structure of Galactic cirrus on the line of sight, and is therefore of no help in separating Galactic and LMC  extinctions.

The relationship between Galactic cirrus and Magellanic sight-lines can be investigated at an even smaller scale with the Palomar visible plates (DSS2 blue plates, available at \emph{Sky-View}).
The plates show, in and around the Magellanic Clouds, nebulosities which 'zoom' into the infrared filaments and reveal the Galactic structures at a $\sim 1''$ scale.
The source of illumination of these nebulosities could be the Magellanic Clouds themselves \citep{o1}, which are the strongest source of visible light in the region.
Most of the Magellanic stars with a bump are observed behind such nebulosities
(see for instance the LMC DSS2 blue images around  Sk-68~155, Sk-68~140, Sk-69~213, Sk-69~270, Sk-69~206). 

The separation between Galactic and Magellanic extinctions seems to be out of reach from up to date observations.
It can not be excluded for example that the $2200\,\rm\AA$ bump observed in some Magellanic directions is due to Galactic cirrus on the lines of sight, while Magellanic extinction laws are always linear.
This would explain why the bump is more frequently observed towards the LMC than towards the SMC, since the cirrus in front of the LMC has an average column density ($E(B-V)\sim 0.075$, $A_V\sim 0.25$, \citet{schlegel98})  about twice that of the cirrus in front of the SMC    ($E(B-V)\sim 0.037$, $A_V\sim 0.12$).
In this case Magellanic extinction laws would resumed to the linear part of the extinction curves.
\section{Conclusion} 
\label{con}
Previous works on Magellanic extinction based on UV extinction curves all find major differences between the characteristics of extinction in the SMC, the LMC, and the Milky Way. 
In contrast, the present analysis shows that there is no proof, up to date, of a different nature of the extinction in the Magellanic Clouds and in the Galaxy, and therefore that there is no reason, from the sole study of extinction curves, to suppose a variation of grain composition between these galaxies.
This analysis  also highlights the important role linear extinction laws play, even in high column density directions.
A generalization of the CCM fit, which still depends upon a single parameter for a given size of the $2200\,\rm\AA$ bump,  was found and applies, in all three galaxies, to all directions except the 'transition' ones where the linear visible reddening law breaks in the far-UV only, and where no $2200\,\rm\AA$ bump is found.

This study first proved (Sects.~\ref{bvn} and \ref{nb}) that, in SMC/LMC low column density directions, Magellanic and Galactic extinction laws are linear and can not be distinguished.
Combined with the  \citet{z01} finding of linear extinction laws in Galactic slightly reddened directions, it implies that linearity over all the visible/UV wavelength range in low column density directions characterizes interstellar extinction in all three galaxies.
As column density increases, linearity breaks first in the far-UV, as an excess of light appears in some sight-lines.
For the highest column densities, extinction is less than expected by the prolongation of the linear visible rise in the UV, and the traditional UV extinction features progressively appear.
There are no fixed thresholds between these transitions in the shape of the extinction curve.
For example, linearity seems to hold at larger $E(B-V)$ in the SMC than in the Milky Way.

These  ostensibly unpredictable behaviors are described by  relationships which, surprisingly, involve a very small number of parameters.
In 1989 the CCM article first proved that a large set of normalized extinction curves is accounted for by a single parameter law\footnote{\emph{The most important result presented here is that the entire mean extinction law, from the near-IR through the optical and IUE-accessible UV, can be well represented by a mean relationship which depends upon a single parameter... the deviations of the observations from the mean relation (e.g., Figs 1 and 2) are impressively small.}' CCM, p.~253}.
This law however is not followed by all extinction curves: linear extinction curves in particular, which characterize the SMC but are also present in the Milky Way, can not be fitted using the CCM law (Sect.~\ref{ccm}).
To embrace an even larger set of extinction curves it is necessary to give up the correspondence, implicit in CCM, between E(B-V) and bump size.
The reddening in any one direction can then be separated into two parts, one following the CCM extinction law, the other a linear one.
This generalization of the CCM fit is given by Eq.~\ref{eq:ccmg}  (Eq.~\ref{eq:ccmgm} for normalized extinction curves).
It applies to all Galactic and Magellanic directions except those  which have no $2200\,\rm\AA$ bump and depart from linearity in the far-UV.

That some directions do not follow the generalized CCM relationship of Sect.~\ref{fit}  should simply  be attributed to its lack of physical foundations.
The large uncertainties on the parameter $R_V$ derived from this fit, its role in  Eq.~\ref{eq:ccmg}  (it affects only the UV part of the fits) question, in my opinion, the physical meaning of $R_V$ derived from either the CCM fit or from  Eqs.~\ref{eq:ccmg} and \ref{eq:ccmgm}.

It is therefore not possible from the data in hand to infer differences between Magellanic and Galactic dusts.
It is also not possible for directions with a linear extinction law to get any precise information on the chemical composition of the dust: even atmospheric aerosols lead to such laws.
Linear extinction is an ubiquitous phenomenon in Nature which provides information on the grain size distribution, but not on the grains' composition.

This study highlights a general coexistence of linear and non-linear components in interstellar extinction laws, which, along with the passage from linearity to CCM extinction curves remain to be understood.
It could be that both types of extinction coexist along some sight-lines, and one could imagine that in the direction of the Magellanic Clouds, the Magellanic extinction law is linear (as it was suggested for the SMC by \citet{prevot84}) while the Galactic law is CCM-like.
This can not be  a complete and satisfactory answer however since it does not help our understanding of the intermediate stage of extinction, where linearity extends in the near-UV only, nor does it explain the role of linear extinction laws in the Galaxy.
The solution should come from a better understanding of these transition directions and will probably imply a re-consideration of the nature of interstellar extinction, in particular  the advent of a fit based on a physical understanding of the phenomena at work.
\section*{Acknowledgments}
This work was supported by a NATO fellowship.

I had a wonderful  working environment during several stays at the Konkoly Observatory of Budapest and wish to thank the members of the Observatory for their help and welcome. I am especially grateful to M\'aria Kun, Lajos Bal\'azs, P\'eter Abrah\'am for having facilitated these stays.

I would also like to thank the anonymous referee for several suggestions which have contributed to the final version of the manuscript.

This research has made use of the SIMBAD and VIZIER databases,
operated at CDS, Strasbourg, France (http://simbad.u-strasbg.fr), of the IUE archives retrieved from the INES Archive Data Server at Vilspa ESA centre (http://ines.laeff.esa.es/cgi-ines/IUEdbsMY), and of NASA's SkyView facility
(http://skyview.gsfc.nasa.gov) located at NASA Goddard Space Flight Center (USA).

{}

\end{document}